# Deep-well ultrafast manipulation of a SQUID flux qubit

**M G Castellano**[1, 4]**, F Chiarello**[1]**, P Carelli**[2]**, C Cosmelli**[3]**, F Mattioli**[1]**, G Torrioli**[1]

[1] Istituto Fotonica e Nanotecnologie - CNR, Roma, Italy
[2] Dip. Ingegneria Elettrica e dell'Informazione, Università dell'Aquila, L'Aquila, Italy
[3] Dip. Fisica, Sapienza Università di Roma, Italy

E-mail: mgcastellano@ifn.cnr.it

**Abstract**. Superconducting devices based on the Josephson effect are effectively used for the implementation of qubits and quantum gates. The manipulation of superconducting qubits is generally performed by using microwave pulses with frequencies from 5 to 15 GHz, obtaining a typical operating frequency from 100MHz to 1GHz. A manipulation based on simple pulses in the absence of microwaves is also possible. In our system a magnetic flux pulse modifies the potential of a double SQUID qubit from a symmetric double well to a single deep well condition. By using this scheme with a Nb/AlOx/Nb system we obtained coherent oscillations with sub-nanosecond period (tunable from 50ps to 200ps), very fast with respect to other manipulating procedures, and with a coherence time up to 10ns, of the order of what obtained with similar devices and technologies but using microwave manipulation. We introduce the ultrafast manipulation presenting experimental results, new issues related to this approach (such as the use of a compensation procedure for cancelling the effect of "slow" fluctuations), and open perspectives, such as the possible use of RSFQ logic for the qubit control.

## 1. Introduction

Superconducting qubits [1] have proven to be very strong candidates for solid state implementation of quantum computing. Artificial atoms, namely two-state quantum systems, can be built using superconducting elements like Josephson junctions (a strongly non-linear element), flux-quantizing loops and so on. According to which degree of freedom is used to monitor the qubit state, the superconducting qubits are named phase [2], flux [3], transmon [4], charge [5] and charge-phase [6] qubits. They can be fabricated with well-known techniques used for integrated circuits. An impressive progress has been made from the very first observation of Rabi oscillations [7], to the first quantum algorithms implemented on two qubits [8].

All these qubit prototypes rely on the use of microwave signals to manipulate and read out the qubits. When one thinks of a system of many qubits, the complexity and the cost of the required instrumentation grows bigger and bigger. In this paper we present an alternative approach, namely controlling a flux qubit by means of fast pulses of magnetic flux, thus avoiding the use of radiofrequency. This method is appealing in the view of full integration of the control electronics on the qubit chip, by using RSFQ logic circuits [9, 10] to provide the pulses and synchronize them. The result would be a fully integrated system, scalable on a large scale, where both qubit and electronics are realized with the same technology.

---

[4] To whom any correspondence should be addressed.

## 2. The double SQUID qubit

The qubit used in this work is based on a double SQUID [11-14] namely a superconducting loop interrupted by a dc-SQUID with much smaller inductance, which behaves as a rf-SQUID whose critical current can be adjusted from outside by applying a magnetic flux.

The schematic of the device is shown in figure 1 a; in the case considered here, the loop inductance is $L$=85pH for the large loop, $l$=6 pH for the small loop, while each Josephson junction has critical current $I_0$=8 µA and capacitance $C$= 0.4 pF. Currents through two different coils couple the control magnetic fluxes $\Phi_x$ (applied to the large loop) and $\Phi_c$ (applied to the small loop); their mutual inductance with the qubit is respectively $M_x$=2.4 pH and $M_c$=6.0 pH. The gradiometric structure of both loops, present in the real device but not shown in the schematic, allows to have small cross coupling between the two fluxes. The relevant degree of freedom for this qubit, in the limit of negligible inductance of the small SQUID ($l$<<$L$), is the magnetic flux $\Phi$ in the large loop; this quantity is read out by a hysteretic dc-SQUID inductively coupled to the main loop, with transforming ratio of 0.01: $\Phi$ is determined by measuring the value of the switching current (from the zero voltage to the running state) for the readout SQUID, a quantity that depends on the coupled magnetic flux [13].

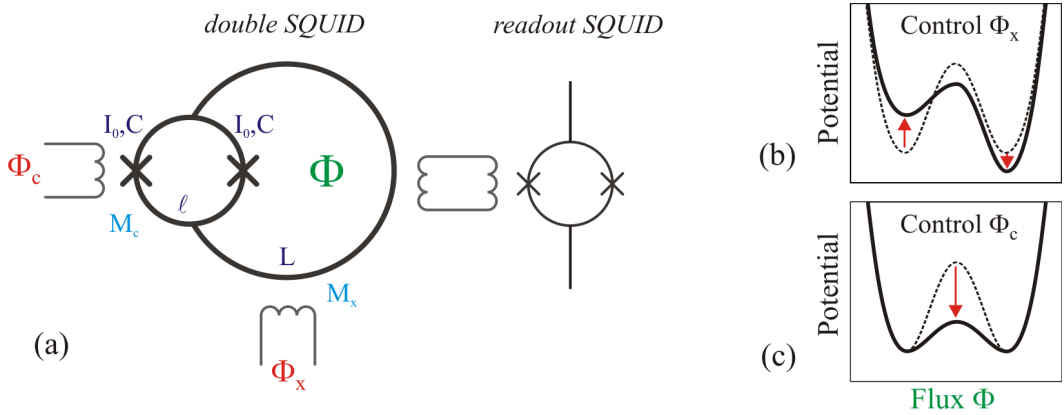

**Figure 1.** (a) Scheme of the double SQUID qubit coupled to the readout SQUID. (b) Effect of the control flux $\Phi_x$ on the potential symmetry. (c) Effect of the control flux $\Phi_c$ on the potential barrier.

By introducing the quantity $\Phi_b=\Phi_0/2\pi$, where $\Phi_0$=h/2e=2.07 10$^{-15}$ Wb is the flux quantum, and expressing the fluxes in reduced units, $\varphi=\Phi/\Phi_b$, $\varphi_x=\Phi_x/\Phi_b$, $\varphi_c=\Phi_c/(2\Phi_b)$, one gets the following expression for the system Hamiltonian :

$$H = \frac{p^2}{2M} + \frac{\Phi_b^2}{L}\left[\frac{1}{2}(\varphi - \varphi_x)^2 - \beta(\varphi_c)\cos\varphi\right], \quad (1)$$

where $p$ is the conjugate momentum, $M = C\Phi_b^2$ is the effective mass and $\beta(\varphi_c)= (2LI_0/\Phi_b)\cos(\varphi_c)$. The potential can be manipulated by using the two control fluxes $\Phi_x$ and $\Phi_c$; in particular, at $\Phi_x=\Phi_0/2$ it takes the shape of a symmetric double well. The energy barrier between the wells can be enhanced or reduced by $\Phi_c$ (figure 1 c), or even made to disappear, so reducing the potential to a single well whose bottom curvature is controlled by $\Phi_c$. By acting on $\Phi_x$, instead, an asymmetry is introduced in the potential (figure 1 b), up to a point where one of the wells disappears. Experimentally, these critical points can be found by preparing the system in one of the two wells and tilting the potential by means of $\Phi_x$ until the initial well becomes unstable and the system switches to the other well (the remaining one). Plotting the positions of such points in the $\Phi_c$-$\Phi_x$ plane, one gets the stability diagram of figure 2: within each lozenge, the potential consists of two wells; outside, the potential is made by a single well [15]. The symmetry axis of the lozenge corresponds to the case of a perfectly symmetric potential. The experimental points (dots) can be fitted (continuous line) to get the estimate of $2LI_0/\Phi_b$; the larger this value, the larger the lozenges. The shape of the lozenges in figure 2 is compatible with a system with $2LI_0/\Phi_b \sim$ 4.5 and $T$=4.2 K, which was the operating temperature for these preliminary measurements. At lower temperature, the width is enhanced because thermal fluctuations are reduced and escape from the metastable well is inhibited. Below the crossover temperature between classical and quantum behaviour, quantum fluctuation mechanisms dominate over thermal escape.

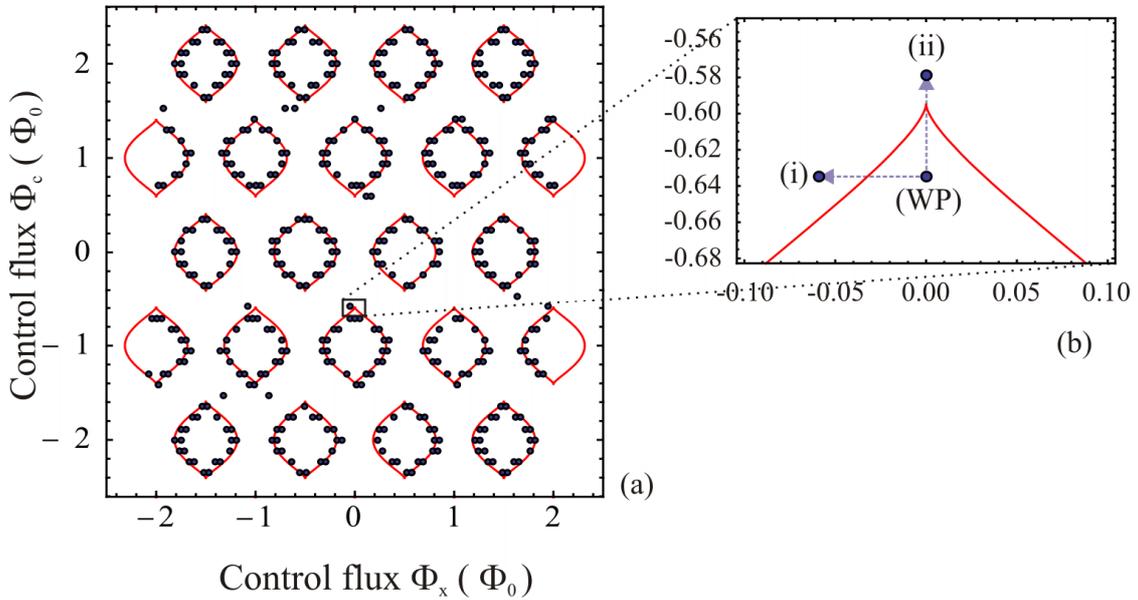

**Figure 2.** (a) Stability diagram of the double-SQUID, in the $\Phi_x$-$\Phi_c$ plane. Solid dots (experimental data, taken at 4.2 K) mark the points where one of the potential wells disappears. Inside each lozenge, potential is a double well, symmetric along the vertical symmetry axis; outside, it is a single well. Continuous line is the fit with the theoretical model. (b) Zoom of the working region at low temperature.

## 3. Operating principle and fast manipulation with pulses

The operating principle [16] of this qubit relies on the interplay between the two types of potential shapes, namely double well and single well, and their base states. Double well and flux basis are used for: initialization of the qubit in a localized state (left or right well); storage of the state by maintaining a high barrier between the wells; state readout. Single well and eigenenergy basis are used for the time evolution of the qubit. The system potential is forced to change from one shape to the other by means of pulses in the control fluxes, with a sequence determined by which function one wants to habilitate. The critical point is how the system passes from one configuration to the other, in other words how the description in terms of flux states and double well is translated in terms of eigenstates of the single well; this depends on how the pulse is applied, either adiabatically or not.

The typical initial working point (WP) lies, with reference to the area depicted in figure 2 (b), close to the tip and to the symmetry axis, where the potential is an almost symmetric double well with a barrier high enough to ensure that escape or tunneling from one well to the other is negligible. To initialize the system in one of the two wells, a pulse on the control flux $\Phi_x$ moves the working point horizontally outside the lozenge to the point (i): here the potential is tilted and only one well is allowed, where the system relaxes. After this pulse, the initial WP is restored and now the system is localized in one well. Likewise, the other well can be populated by reversing the sign of the flux pulse.

In the next step, a pulse on the control flux $\Phi_c$ with top value $\Phi_c^{top}$ moves the WP vertically beyond the tip of the lozenge to point (ii), driving the system in the region where the potential is a single well (figure 3a). The goal of this modification is to populate equally the 0 and 1 energy levels in the single well and use free oscillations of the relative phase between them as the basic operating mechanism of the qubit. The larger the flux pulse, the deeper is the well and the higher is the frequency of oscillations. In contrast with the initialization procedure, during this step the potential symmetry is not affected, except for a possible cross-coupling between the large loop and the small loop of the double-SQUID. Figure 3b shows how the energy level structure changes, passing from a doublet structure (typical of a symmetric double well) to that of a harmonic oscillator. In this figure are plotted the six lowest-lying levels (referred to the ground state for convenience) for a realistic case; on the left, doublets are so close that they are not distinguishable and open up when the barrier disappears.

Reaching the equi-populated condition is crucially depending on the rising rate of the $\Phi_c$ pulse; a slow adiabatic control-flux pulse will not create the desired linear superposition between the |0> and |1> eigenstates needed to demonstrate Larmor oscillations in our qubit. To put it clearer, let's consider first an ideal case of perfectly symmetric potential; in this case, the starting flux eigenfunctions (left or right) do coincide exactly with the superposition of the |0> and |1> eigenstates, so that even an adiabatic manipulation would bring to the desired result. On the contrary, if there is even a tiny asymmetry like in any realistic system, this mechanism does not hold. For example, a tilting of just 0.1 n$\Phi_0$ is enough to make flux and energy eigenstates coincident. In this case a slow (adiabatic) transformation cannot change the states occupation, but a fast (non-adiabatic) process with pulse risetime of the order of nanoseconds can equally populate the first two energy levels.

Figure 3c shows the results of a simulation for the realistic case of a system deviating from perfect symmetry on $\Phi_x$ by 0.1$\mu\Phi_0$ and subjected to a $\Phi_c$ control flux linearly increasing in time, with a rate $w=5m\Phi_0/ns$. In this case the qubit is initially prepared in its ground state, coincident with the left well. During the manipulation, the populations remain unvaried until the barrier is strongly reduced ("portal region") and there is a transition that leads to an equal population of the final levels |0> and |1>.

The pulse risetime must be fast enough to allow for this behavior, but not too fast in order to avoid the population of levels above the first two. Fortunately, there exist a range of pulse risetimes where both these conditions can be met, thanks to the presence of an energy gap (figure 3b) that separates the first doublet from upper ones. Figure 3d-f show the simulated evolution of the populations as in the case of figure 3c but with increasingly faster pulses; 3c and partially 3d are two examples of acceptable behavior, while situations like 3e and 3f must be avoided.

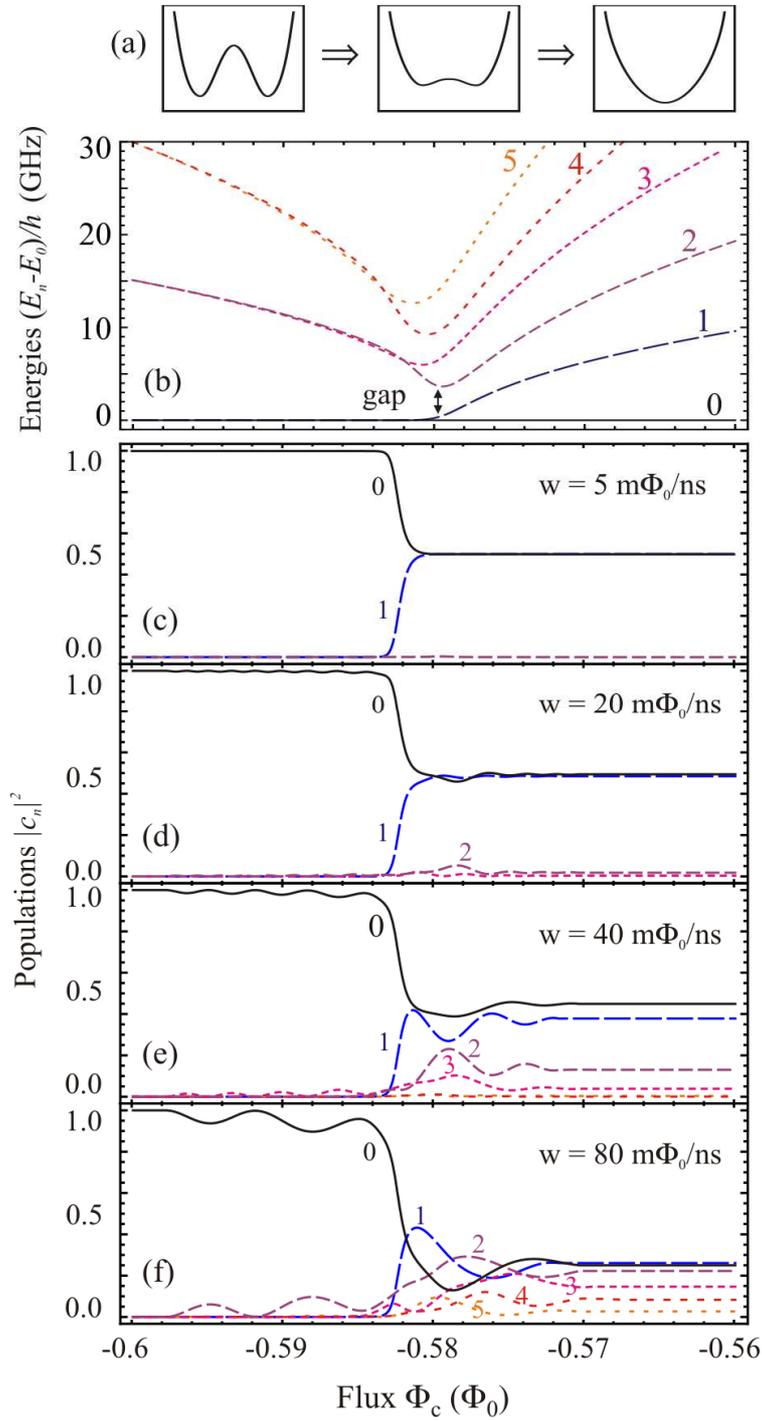

**Figure 3.** (a) Schematic of the potential change considered in the simulation. (b) Modification of the lowest-lying six energy levels, referred to the ground level. (c-f) Evolution of the state populations for different rising rates of the $\Phi_c$ control pulse (simulation for a deviation of $0.1\mu\Phi_0$ from the perfect symmetry and $\Phi_c$ linearly increasing with time). Only cases c and d are suitable for qubit operation, while cases e and f, with faster $\Phi_c$ risetime, show the excitation of unwanted levels and are to be avoided.

After this step, the $\Phi_c$ value is maintained constant for a while. During this uptime $\Delta t$, the system, which is now in the single well condition with only the first two levels equally populated, is let to evolve freely. This evolution does not involve transitions between the levels (except for relaxation and possible excitation due to external noise) but it affects only relative phases. During this time the system is quite protected from external disturbances because the single well potential shape is only weakly responding to the bias flux parameters.

Successively, during the fall time of the control pulse, the barrier is raised again, transforming back the potential into a double well. The relative phase gained during the uptime is projected into the population of the left/right flux state, with a process that is the opposite of what occurred during the risetime. The manipulation sequence produces periodic oscillations of the left/right population as a function of $\Delta t$, with an oscillation frequency depending on the level spacing in the single well condition.

We point out that our qubit and the manipulation scheme present several similarities with the IBM qubit developed by R. H. Koch [17]. Although the physical implementation is different (electrical scheme and materials), both qubits rely on the manipulation of a double well potential by means of magnetic flux pulses that reduce the barrier height and modify the energy level structure, in particular the fundamental doublet used for the computation. For both qubits, the way this change is performed (adiabatically or non-adiabatically) is essential. The main difference is how the system is stabilized against the extreme instability due to the exponential dependence of the doublet spacing on the barrier height. The IBM group couples the qubit to a transmission line that is used as an external oscillator, while in our case the oscillator is realized by the qubit potential itself, in its single-well embodiment: this adds the possibility of frequency tuning and higher frequency values.

## 4. Experimental setup and results

The qubit measured in this work was fabricated by Hypres [18] with a Nb/AlOx/Nb trilayer process with 100 A/cm$^2$ critical current density and SiO$_2$ as dielectric insulator. The nominal qubit parameters are given in section 2. The chip is included in a OFHC copper case that is thermally anchored to the mixing chamber of a $^3$He-$^4$He dilution refrigerator [19]. The electrical leads are made by phosphor-bronze wires, filtered by CLC filters at room temperature and by RCR filters on the refrigerator still, and by Thermocoax [20] down to the lowest temperature stage (30 mK); the overall cutoff frequency is about 100 kHz. The coil for $\Phi_c$ is fed also by a "fast" line, a 50 $\Omega$ coaxial line made by Nb; thermalization and filtering are achieved by means of 20dB attenuators placed on the 1K pot and at the lowest temperature stage. At low temperature, because of the variation of the physical parameters of the cables, a perfect matching between the last section of the coax line and the on-chip circuit is not guaranteed. Fast and slow lines for the $\Phi_c$ coil are joined on the chip holder. Flux pulses are provided by an Agilent 81130A pulse/data generator and have fixed nominal risetime of 0.5 ns.

The main result of our measurements is the observation of the coherent free oscillations of the flux state populations as a function of the $\Phi_c$ pulse duration $\Delta t$ for different values of $\Phi_c^{top}$ (figure 4). Each experimental point in any curve is obtained by repeating many times (100-1000 times) the manipulation sequence described in the previous paragraph, in order to estimate the probability to find the system in the left/right well.

An online correction of the working point has been used in order to reduce the effect of slow fluctuations of the bias flux, and will be described in detail in par. 4.3.

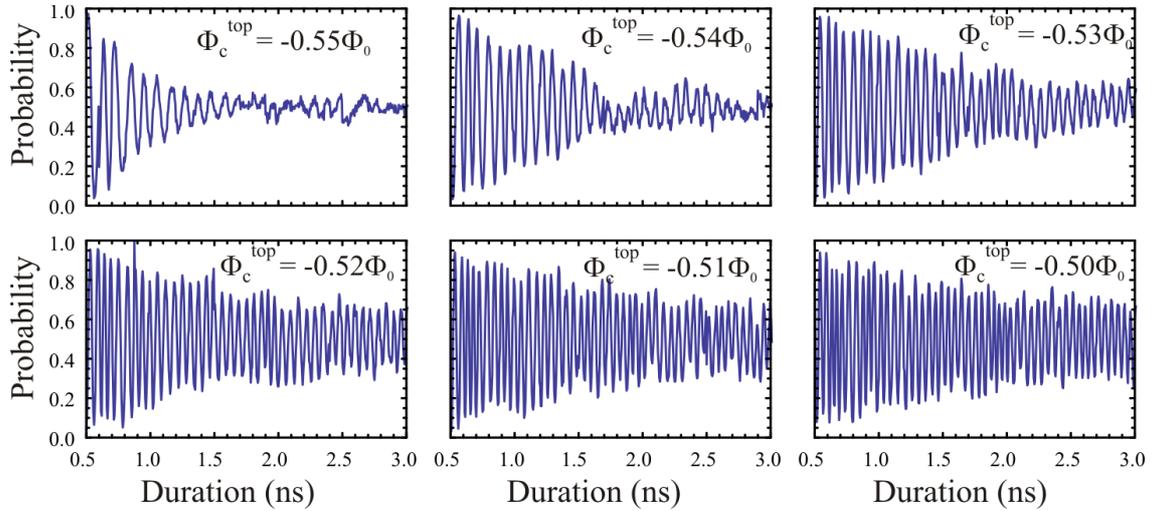

**Figure 4.** Oscillations taken at 30 mK with increasing depth of the potential single well. Oscillation frequency increases from 10.5 to 26.7 GHz.

*4.1. Frequency tunability*

By increasing the height of the $\Phi_c$ pulse, the working point at which the qubit undergoes free evolution moves further from the lozenge tip, corresponding to deeper single well potential. As a consequence, the distance between the first two levels grows larger, as well as the oscillation frequency (see the different plots of figure 4). This mechanism allows tuning of the qubit operating frequency from a few GHz up to more than 20 GHz (figure 5).

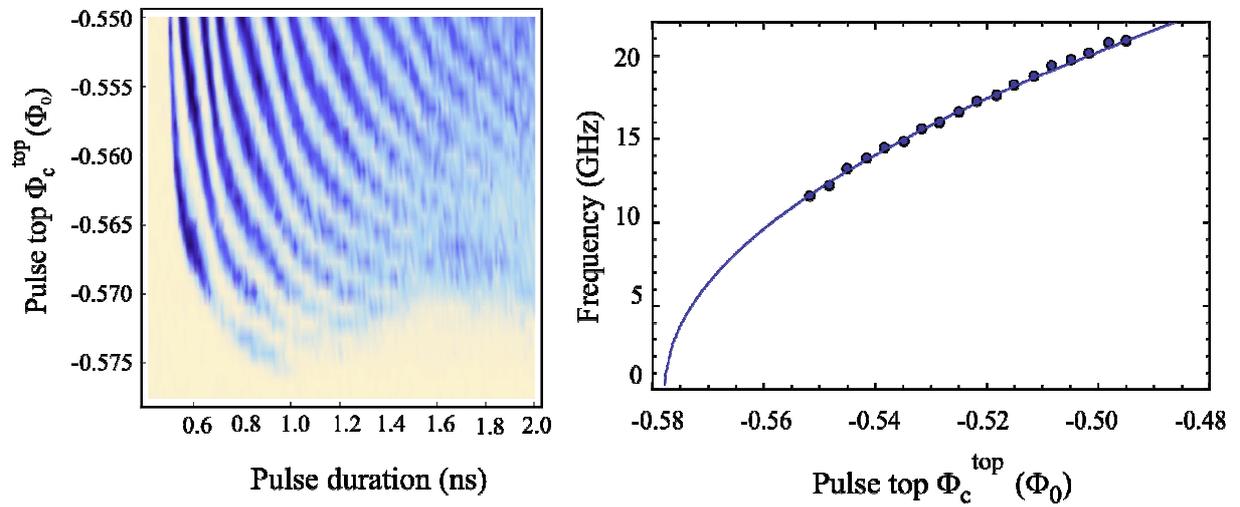

**Figure 5.** Left: density plot of the probability oscillations (z-axis) vs. duration (x-axis) and height (y-axis) of the $\Phi_c$ pulse. Right: oscillation frequency in GHz as a function of the pulse height ($\Phi_c^{top}$). continuous line is the theoretical fit.

Going back to figure 4, we see that, besides frequency, both the decay shape and the decay time depend on the value of $\Phi_c^{top}$. For smaller pulses we observe a typical exponential decay, as expected by considering a simple Ohmic model for the noise. For larger pulses the decay is no more exponential, as can be expected if the low frequency noise component is dominant [21], and the coherence time is longer, leading to oscillations that remain visible for a time up to 10 ns (figure 6).

Decoherence times of this order of magnitude are usual for Nb/AlOx/Nb qubits with $SiO_2$ dielectric, as shown in the literature mainly for phase qubits. For devices manufactured by Hypres like ours, Lisenfeld [22] measured an energy relaxation time $T_1$ =1.9 ns and a decay time for Rabi oscillations T´= 3-5 ns. The Maryland group, in several subsequent experiments on Hypres qubits, reported $T_1$=4ns [23], spectroscopic coherence time $T_2^*$ = 0.9 ns and estimated $T_1$=7 ns [24], T´= 9-15 ns and spectroscopic coherence time $T_2^*$ = 3-8 ns [25]. A twin of the qubit used in this paper, operated as a phase qubit with microwave excitation, showed T´= 1.5-2 ns and $T_1$ =1.37 ns [26].

Even though these figures are compatible with our experimental results, care should be used in a direct comparison. In fact, we remark that the usual scheme for decoherence times in microwave forced qubits does not apply to our manipulation scheme, where a free evolution of the system is preceded and followed by two rapid modifications of the potential shape in a non adiabatic limit. In-depth studies of the various sources of dephasing and relaxation for our qubit have yet to be performed.

Figure 6 shows one of the best oscillations, at a frequency of 16.6 GHz; the experimental points are fitted by a continuous line (green online) as a guide for the eyes, while a dotted line (red online) marks the fit of the envelope to highlight the amplitude decay. This figure emphasize one of the advantages of our particular operating mode that, thanks to a very high oscillation frequency, allows to have many oscillation periods and perform several quantum operations even within a short decay time.

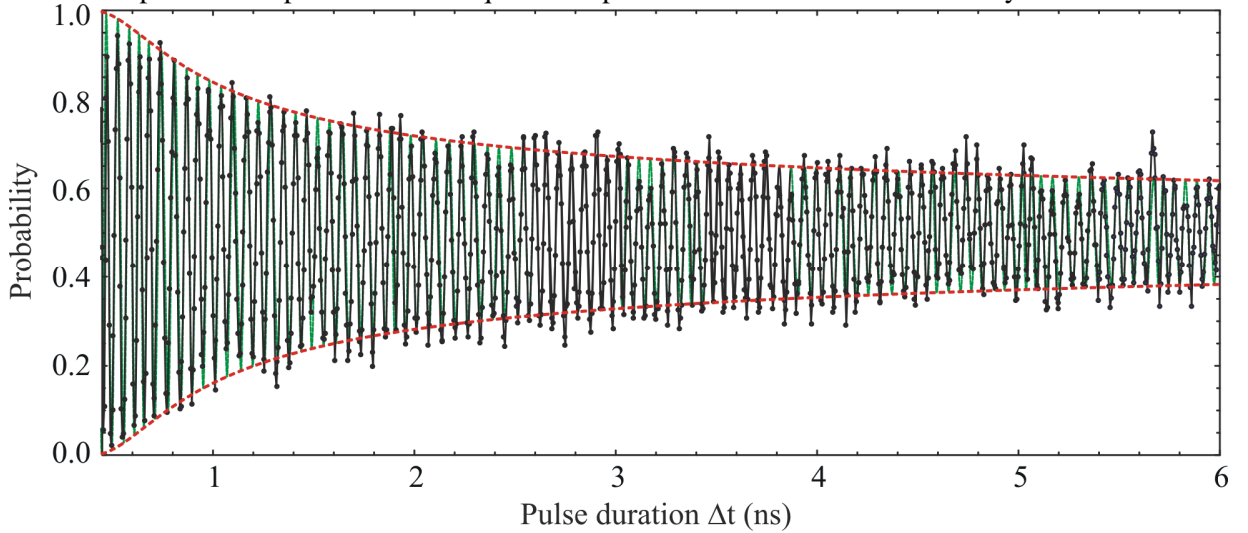

**Figure 6.** One of the best experimental curves showing coherent oscillations at a frequency of 16.6 GHz. The fit of the envelope is marked by a dotted line (red online).

*4.2. Time-frequency analysis*
By analyzing the time domain oscillation curves, one finds that frequency is not constant but changes along the time axis (i.e., duration of the $\Phi_c$ pulse): the oscillation is non-stationary. We then perform a time-frequency analysis [27] by sectioning the time domain data in parts (each containing enough periods), and analyzing each part separately, finding which frequency fits the data at which time. Figure 7 shows the result for two distinct experimental setups, which differ for the circuit that joins fast and slow lines on chip, at $\Phi_c$ terminals. In setup 1, fast and slow lines are simply joined into a single line just before the chip, while in setup 2 a 50 Ω resistance is inserted in the fast line and a 50 nH inductance is inserted in the slow line just before the joining. We recall that in both cases 50Ω matching to the feeding coax at low temperature may not be guaranteed. For each set of data, different lines correspond to different heights of the $\Phi_c$ pulse, namely different depths of the single well potential. Ideally, we would expect that the curves are parallel horizontal lines, corresponding to the constant oscillation frequency in that particular potential shape; instead, we find an additional modulation whose shape is the same for the different curves of each set. We attribute this effect to the not perfect shape of the $\Phi_c$ pulse when it reaches the chip, which is not a trapezoidal pulse with flat top but may present overshoots and ripples because of reflections along the line. Setup 2, with the introduction of additional inductances across the coil for $\Phi_c$, is affected from this modulation more than setup 1; further improvements in the matching circuit should reduce the problem.

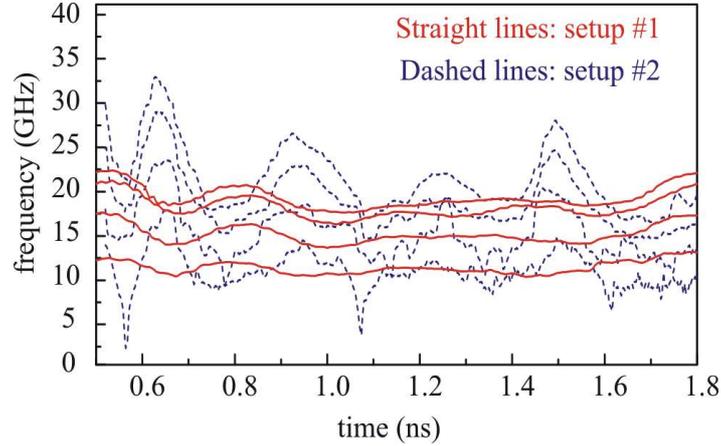

**Figure 7.** Time-frequency analysis for two different setups. For each setup, different curves refer to different depths of the single well potential, determined by the height of the control pulse $\Phi_c$. Higher curves correspond to deeper wells (note that the shape for each setup is the same). In an ideal system, curves should be horizontal lines, corresponding to constant frequency values. Deviation from this behavior is more evident in setup 2 than in setup 1.

### 4.3. Compensation procedure

All the oscillation curves shown until now have been obtained by implementing a compensation procedure that allows reducing the effect of slow fluctuations of the qubit working point.

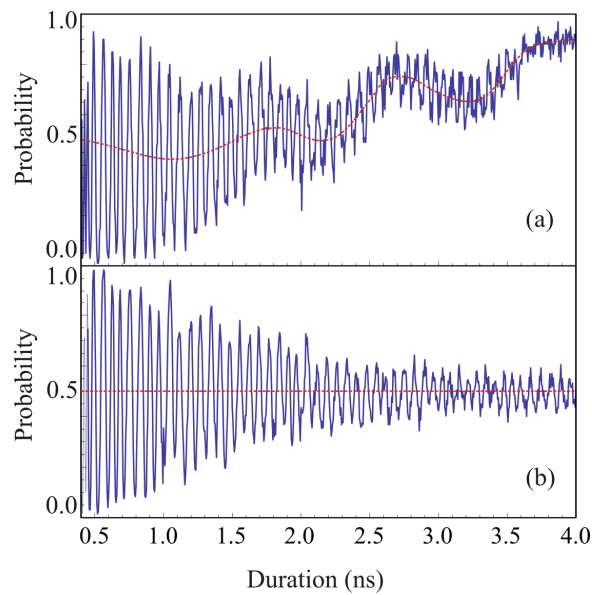

**Figure 8.** (a) Measured oscillations in the absence of corrections,

showing the fluctuation of the middle point. (b) The same oscillations by adopting the compensation procedure of the bias flux $\Phi_x$.

Figure 8 (a) shows the shape of an oscillation of the occupation probability recorded without correction. We note that after an initial part where the oscillation is centered about the value of 50%, this middle point starts wandering up and down, while the oscillation is still on: this behavior can be explained by the value of the bias flux $\Phi_x$ not being stable. By repeating the measurement in different conditions (different pulse top and/or base), we see that the middle point does not move at random, but it follows a repeatable pattern. We then attribute this effect to an unwanted coupling between the coil for $\Phi_c$ and that for $\Phi_x$, such that the pulse on $\Phi_c$ excites a resonant mode on the $\Phi_x$ line circuitry, and moves the middle point away from 50% probability. While of course it is desirable to get rid of the cross talk by intervening on the chip layout and circuitry, it is nonetheless possible to cope with it by compensating the effect on $\Phi_x$ during the measurement. The measurement time is chopped up into several segments, each including a few oscillation periods; for each segment, the acquisition system evaluates the middle value of the occupation probability and changes the dc value of the bias flux $\Phi_x$ until the equilibrium point is again at 50%. With this procedure, for each value of the pulse duration a different value of $\Phi_x$ is supplied. The flux excursion of the correction signal on $\Phi_x$ is at most 3 m$\Phi_0$, a value large enough to move the oscillation centre and to affect the coherence time, but not sufficient to destroy the mechanism generating the oscillations; the shape is reproducible, as expected from a deterministic signal.

The result of the correction is shown in figure 8 (b): the oscillation is now centered about the 50% line. We remark that it is possible to apply this procedure only thanks to the fact that the fluctuation of $\Phi_x$ is much slower than the oscillation frequency.

## 5. Conclusions

The reported measurements show how the qubit can be manipulated just using fast pulses of magnetic flux. The complete qubit manipulation requires also the capability to control the relative phase of its coherent superposition. This can be achieved by exploiting a slight potential unbalance for a short time in order to induce a controlled phase difference. In all cases, it is necessary to work with pulses of magnetic flux with risetime in the order of nanosecond, which should eventually be synchronized by a fast clock. These requirements naturally call for circuits realized with Rapid Single Flux Quantum (RSFQ) logic, based on the processing of individual flux quanta [9].

RSFQ circuits are naturally suited for combining with superconducting qubits because of speed, scalability, compatibility with the qubit fabrication process and low temperature environment. One RSFQ characteristic potentially fatal for qubits is the need of resistors for biasing purposes and for getting shunted Josephson junctions, which can induce decoherence in the qubit circuit just because of their presence in the circuit. Besides, heating due to Joule effect is significant at very low temperature, in spite of the small energy cost of RSFQ circuits, because it can produce hot quasiparticles that again are detrimental for the qubit operation. However, it is possible to remove such obstacles by using different structures and ideas [10] and/or several precautions. For the thermal problems, a specially designed process can reduce power dissipation to just 25 pW for junction, while the use of copper cooling-fins improves refrigeration of the resistive shunts at temperatures in the mK range [28, 29]. As regards the effect of dissipation on qubit decoherence, it can be shown that, with the enhanced fabrication process, it is possible to design circuits such that this issue is overcome [30, 31]. Another difficulty is that the risetime of RSFQ pulses is too high and it would induce excitation to non-computational states in the qubit. Even in this case, however, it is possible to use on-chip filters to slow down pulse risetime or to develop an RSFQ pulse generator made by a series of individual pulses, designed so as to achieve a risetime within the desired range. First attempts in coupling RSFQ circuits to qubits gave encouraging results, although still in the incoherent regime [32].

The possibility of an all-integrated chip with both qubits and electronics makes this type of qubit very attractive for future implementations.

**Acknowledgments**
We are grateful to S. Poletto for his comments and careful reading of the manuscript. This work has been supported by CNR under RSTL project.